\documentclass[aps,prb,groupedaddress,showpacs,twocolumn,superscriptaddress,longbibliography,10pt]{revtex4-2}

\usepackage{graphicx}
\usepackage{mathtools}
\usepackage{amsmath}
\usepackage{amssymb}
\usepackage{bbm}
\usepackage{bm}
\usepackage{cancel}
\usepackage{xspace}
\usepackage{braket}
\usepackage{xcolor}   
\usepackage{calrsfs}
\usepackage{comment}
\usepackage{placeins}
\usepackage{units} 
 
\usepackage[export]{adjustbox}

\usepackage[colorlinks=true]{hyperref} 
  
\DeclareMathAlphabet{\pazocal}{OMS}{zplm}{m}{n} 

\newcommand{\ee}{\mathrm{e}}  
\DeclareMathOperator*{\ii}{i} 
\newcommand*\dd{\mathop{}\!\mathrm{d}}

\renewcommand{\vec}[1]{\bm{#1}} 
\newcommand{\mat}[1]{\bm{#1}} 
\newcommand{\kel}[1]{\underline{#1}} 

\newcommand{\locGFRinv}[1]{\left[ \mathbf{G}^{-1} \right]^{\text{R}}_{#1}}
\newcommand{\locGFKinv}[1]{\left[ \mathbf{G}^{-1} \right]^{\text{K}}_{#1}}
\newcommand{\locRR}[1]{\mathbf{R}^{\text{R}}_{#1}}
\newcommand{\locRK}[1]{\mathbf{R}^{\text{K}}_{#1}}
\newcommand{\locRA}[1]{\mathbf{R}^{\text{A}}_{#1}}
\newcommand{\locLR}[1]{\mathbf{L}^{\text{R}}_{#1}}
\newcommand{\locLK}[1]{\mathbf{L}^{\text{K}}_{#1}}
\newcommand{\locLA}[1]{\mathbf{L}^{\text{A}}_{#1}}

\newcommand{\locGFA}[1]{\mathbf{G}^{\text{A}}_{#1}}
\newcommand{\locGFK}[1]{\mathbf{G}^{\text{K}}_{#1}}

\newcommand{\fig}[1]{Fig.~\ref{#1}}

\DeclareMathOperator*{\rre}{Re}
\DeclareMathOperator*{\iim}{Im}
\DeclareMathOperator*{\Tr}{Tr}

\newcommand{\aamp}{\mathcal{A}}

\definecolor{orange}{RGB}{252,77,6}
\definecolor{brown}{RGB}{200,127,50}
\definecolor{blue}{RGB}{0,80,255}

\definecolor{ao(english)}{rgb}{0.0, 0.5, 0.0}

\begin{document}

\title{Photodriven Mott insulating heterostructures: \\ A steady-state study of impact ionization processes} 

\author{Paolo Gazzaneo}
\email[]{paolo.gazzaneo@tugraz.at}
\affiliation{Institute of Theoretical and Computational Physics, Graz University of Technology, 8010 Graz, Austria}
\author{Daniel Werner}
\affiliation{Institute of Theoretical and Computational Physics, Graz University of Technology, 8010 Graz, Austria}
\author{Tommaso Maria Mazzocchi}
\affiliation{Institute of Theoretical and Computational Physics, Graz University of Technology, 8010 Graz, Austria}
\author{Enrico Arrigoni}
\email[]{arrigoni@tugraz.at}
\affiliation{Institute of Theoretical and Computational Physics, Graz University of Technology, 8010 Graz, Austria}

\date{\today}


\begin{abstract}    
   
We investigate the photocurrent and spectral features in a simplified model of a Mott photovoltaic system consisting of a multilayered insulating heterostructure. The central correlated region is coupled to two metallic leads kept at different chemical potentials. A periodic drive applied to the correlated region produces excited doublons and holons across the Mott gap which are then separated
by a potential gradient, which mimics the polarization-induced electric field present in oxyde heterostructures. The nonequilibrium Floquet steady-state is addressed by means of dynamical mean-field theory and its Floquet extension, while the so-called auxiliary master equation approach is employed as impurity solver. We find that impact ionization, identified by a kink in the photocurrent as function of the driving frequency, becomes significant and is generally favoured by weak, narrow-band hybridizations to the leads beyond a certain strength of the driving field. On the other hand, in the case of a direct coupling to metallic leads with a flat band, we observe a drastic reduction of impact ionization and of the photocurrent itself. 
  
\end{abstract} 
 



       
\maketitle

\section{Introduction}\label{sec:intro}

The current climate crisis calls
 for innovation in new sustainable energy production, such as the field of photovoltaic cells. While conventional semiconductors and organic solar cells are currently employed in most applications, correlated Mott insulators have been proposed as interesting alternatives. In particular, it has been suggested that the Mott gap might be used to convert electromagnetic radiation into electric energy ~\cite{mano.10,li.ch.13,gu.gu.13,co.ma.14,wa.li.15} with an efficiency beyond the Shockley-Queisser limit~\cite{sh.qu.61}, due to so-called impact ionization (II)~\cite{mano.10,co.ma.14} processes. 
Beyond theoretical conjectures, II, and more generally multiple exciton generation (MEG) processes, have been detected via photocurrent spectroscopy and pump-probe experiments in VO$_2$~\cite{ho.bi.16,sa.kh.23} and in quantum dots ~\cite{fr.an.06,wa.mc.13,wa.bo.17}.

In particular, oxide heterostructures based on LaVO$_3$/SrTiO$_3$ have been identified as possible candidates~\cite{as.bl.13} for Mott photovoltaics, even though some issues, such as the low mobility of the carriers~\cite{wa.li.15,je.re.18}, may prevent the use of such compounds as efficient solar cells. Nevertheless, recent years have witnessed an increasing effort towards the growth and characterization of LaVO$_3$ thin films, see, e.g.~\cite{zh.xi.21}. 
The characterization of Mott-based photovoltaic devices and the underlying physics governing fast carrier multiplication processes are subjects of broad interest, with potential applications across various domains. Indeed, numerous theoretical investigations have delved into diverse facets of II, see, e.g.~\cite{co.ma.14,ec.we.11,ec.we.13,we.he.14,pe.be.19,so.do.18,ka.wo.20,mano.19,ma.ev.22,wa.wa.22}.

Heterostructures with metal oxides compounds like LaVO$_3$ exhibit features which are typical of Mott insulators, due to the strong local Coulomb interactions of the electrons in the outer \emph{d}-shells of Vanadium atoms, and therefore they may be regarded as prototypes for real-space dynamical mean-field theory (DMFT)~\cite{free.04, zl.fr.17, ma.am.15, okam.07, okam.08, ec.we.13, ec.we.14, ti.do.16, ti.so.18} studies. The Floquet generalization of this formalism, which allows to include external periodic drivings, provides an important benchmark for nonequilibrium real-time DMFT for multilayer setups at long times, which are computationally costly to address~\cite{ec.we.13,pe.be.19}.

\begin{figure}[b] 
\includegraphics[width=0.8\linewidth]{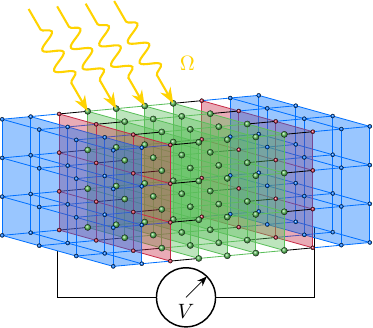}
\caption{System under investigation. The central region consisting of $\text{L}=4$ correlated layers, in green, is irradiated with a periodic, monochromatic, light with frequency $\Omega$. Two uncorrelated interfaces acting as contacts, in red, separate the correlated layers from the metallic leads, in blue, which are kept at different chemical potentials, and model the collectors for the charges created by the external driving $\Omega$.}
\label{fig:setup_voltmeter_final}
\end{figure}

In the present manuscript we characterize a prototype of solar cell based on the heterostructure schematically depicted in Fig.~\ref{fig:setup_voltmeter_final} in terms of the occurrence of II. The system under investigation consists of $\text{L}=4$ Mott insulating layers located between two wide-band metallic leads which act as charge collectors to harvest the energy. The Mott region is separated from the metallic leads by two narrow-band layers with energies aligned with the upper and lower Hubbard bands as in Fig.~\ref{fig:energy_setup_final}, which represent the oxide contact layers. An external time-periodic electric field induces a Floquet steady state with an average current flowing from the left to the right lead, against the potential energy established by the leads' chemical potentials. Due to the polar interfaces located between the insulating layers~\cite{as.bl.13}, an internal electric field builds up in the heterostructure, acting as a separation mechanism for the electron-hole (e-h) pairs produced by the periodic driving.
     
Our goal is the characterization of this Mott-based photovoltaic setup in a nonequilibrium steady-state (NESS) in terms of the occurrence of II and we do so by analyzing some of the most important observables such as the steady-state current. In addition to that, we investigate how the electric field amplitude of the external driving affects II. We then discuss the situation in which the Mott region is directly connected to wide-band metallic leads without intermediate narrow-band contacts, and address II in this case. The strong-coupling impurity solver we use allows us to obtain accurate results directly in the NESS. The downside is that we are restricted to a single-band model, so that we cannot address Hund's coupling effects as in Ref.~\cite{pe.be.19}. Also the effects of local antiferromagnetic correlations as discussed in Ref.~\cite{ec.we.14} are beyond the scope of our present study.

Our results indicate that II, which we link to the presence of a sharp increase (``kink'') in the photocurrent for driving frequency strengths around twice the gap, plays an important role when the correlated region is connected to narrow-band contacts.
This sudden increase in the photocurrent is generally supported by a corresponding kink in the double occupation and its fluctuation around its mean value (see Eq.~\eqref{eq:dev_double_occ} for details) as well as by a non-negligible occupation of the upper bands for driving frequencies compatible with II.
Upon decreasing the electric field amplitude $E_0$, our results suggest that II is accompanied by a change of behavior of the photocurrent from quadratic to linear as function of $E_0$. Factors that hinder II turn out to be (i) a stronger hybridization to the leads, as we already observed in the case of a single layer~\cite{ga.ma.22}, as well as (ii) a direct connection to wide-band metallic leads without intermediate narrow-band contacts.
  
The manuscript is organized as follows: In Sec.~\ref{sec:Model} we introduce the setup under investigation and its Hamiltonian. In Sec.~\ref{sec:Method_formalism} we discuss the nonequilibrium Green's function formalism and the mathematical tools hereby employed. Results are discussed in Sec.~\ref{sec:results} while Sec.~\ref{sec:conclusions} is left for conclusions and final remarks.
    
\section{Model}\label{sec:Model}

We consider a system made of $\text{L}$ correlated layers arranged along the $z$-axis, i.e. $z\in \left\{ 1,...,\text{L} \right\}$, translationally invariant in the $xy$ plane, attached to two metallic leads ($\rho=\text{l},\text{r}$) as shown in \fig{fig:setup_voltmeter_final}. 

The Hamiltonian of the system reads
\begin{equation}
\label{eq:Hamiltonian}
\begin{split} 
\hat{H}(t) & = - \sum_{z, \langle {\bf r},{\bf r}'\rangle, \sigma}t_{z}(t) \hat{c}_{z,{\bf r},\sigma}^\dagger \hat{c}_{z,{\bf r}',\sigma}
- \sum_{\langle z, z'\rangle, {\bf r}, \sigma} t_{zz'} \hat{c}_{z,{\bf r},\sigma}^\dagger \hat{c}_{z',{\bf r},\sigma}
\\ & + \sum_{z,{\bf r}} U_{z} \hat{n}_{z,{\bf r},\uparrow} \hat{n}_{z,{\bf r},\downarrow} + \sum_{z,{\bf r},\sigma} \varepsilon_z^{(0)} \hat{n}_{z,{\bf r},\sigma} + \hat{H}_{\text{leads}}.
\end{split}
\end{equation} 
The operator $\hat{c}_{z,{\bf r},\sigma}^\dagger$ ($\hat{c}_{z,{\bf r},\sigma}$) creates (annihilates) an electron on site ${\bf r}=(x,y)$  of  layer $z$ with spin $\sigma= \lbrace \uparrow,\downarrow \rbrace$. $\hat{n}_{z, {\bf r}, \sigma}=\hat{c}_{z, {\bf r}, \sigma}^\dagger \hat{c}_{z, {\bf r}, \sigma}^{\phantom\dagger}$ denotes the particle number operator on the correlated layer $z$. Here the brakets $\langle z, z'\rangle$ denote neighboring layers along the $z$-axis while $\langle {\bf r}^{\phantom\dagger},{\bf r}'\rangle$ is referred to neighboring sites belonging to the same layer. 

The time-dependence in the Hamiltonian~\eqref{eq:Hamiltonian} is due to the time-periodic, homogeneous and monochromatic, electric field with frequency $\Omega$ and enters via the Peierls substitution~\cite{peie.33} in the intralayer hopping as
\begin{equation}
\label{eq:peierls} 
t_{z}(t) = t_{z} \ e^{-\ii \frac{q}{\hbar} \left( \vec{r} - \vec{r}' \right) \cdot \vec{A}(t)}.
\end{equation}
In Eq.~\eqref{eq:peierls} $\vec{A}$(t) is the time-dependent vector potential, $\hbar$ Planck's constant and $q$ the charge of the electron. We choose $\vec{A}(t)= A(t)\vec{e}_{0}$ to lie along the lattice body diagonal of a hyper-cubic lattice $\vec{e}_{0}=(1,1,\dots,1)$, with $A(t)=\frac{\hbar}{qa}\aamp\sin(\Omega t)$ and $\aamp=-\frac{qE_0a}{\hbar\Omega}$, where $E_0$ is the electric field amplitude, and $a$ the lattice spacing~\cite{ts.ok.08,mu.we.18}.
In the temporal gauge the electric field is then given by $\vec{E}= -\partial_{t}\vec{A}(t) = E_0 \cos(\Omega t) \vec{e}_{0}$.
 
The second term in Eq.~\eqref{eq:Hamiltonian} accounts for electron hopping processes in between layers and is described by the nearest-neighbor amplitude $t_{zz'}$. The third term introduces the local onsite Hubbard interaction $U_z$ and the fourth one describes the onsite energies $\varepsilon_z^{(0)}$, the expression of which will be given in Sec.~\ref{sec:Dyson_equation}. $\hat{H}_{\text{leads}}$ represents the Hamiltonian of the metallic leads, the details of which will be also presented in Sec.~\ref{sec:Dyson_equation}. 

In this manuscript we consider a uniform Hubbard interaction $U_{z}=U$. For the correlated region, we set $t_{z}\equiv t_{\parallel}$ and $t_{zz'}= t_{\perp}$ with $t_{l,1}=t_{L,r}=v_{\rho}$ so that the hybridization between the correlated region and the leads is the same on both sides of the heterostructure. 

As already mentioned, the \emph{xy}-plane is modeled as a $d$-dimensional lattice which, in the limit $d \rightarrow \infty$ introduces the usual rescaling $t_{\parallel}=t_{\parallel}^{\ast}/(2\sqrt{d})$~\cite{tu.fr.05}. In this framework, sums over the electron crystal momentum ${{\bf k}}$ transform as $\sum_{{{\bf k}}} \chi(\omega,{{\bf k}}) \rightarrow \int \dd \epsilon \int \dd \overline{\epsilon} \ \rho(\epsilon,\overline{\epsilon}) \chi(\omega;\epsilon,\overline{\epsilon})$, $\rho(\epsilon,\overline{\epsilon}) = (1/\pi t_{\parallel}^{\ast 2}) \exp[-( \epsilon^{2} + \overline{\epsilon}^{2})/t_{\parallel}^{\ast 2}]$ being the joint density of states~\cite{ts.ok.08}, where 
\begin{align}\label{eq:d-dim_crystal_dep}
\begin{split}  
\epsilon & = -2t_{\parallel} \sum_{i=1}^{d} \cos(k_i a), \\  
\overline{\epsilon}& = -2t_{\parallel} \sum_{i=1}^{d} \sin(k_i a).
\end{split}  
\end{align}
Throughout the whole manuscript we choose our units so that $\hbar = k_{\text{B}} = a = 1 = -q$ and $t_{\parallel}^{\ast}/2=1$ as unit of energy.

\section{Method and formalism}\label{sec:Method_formalism}   

We describe the periodic NESS using the Floquet generalization of the nonequilibrium Green's function (GF) formalism~\cite{ts.ok.08,sc.mo.02u,jo.fr.08}. In this work a Floquet-represented matrix is denoted as either $X_{mn}$ or $\bf{X}$, while underline stands for the \emph{Keldysh} structure
\begin{equation}\label{eq:Keld-structure}
\underline{{\bf G}} \equiv 
\begin{pmatrix}
{\bf G}^{\text{R}} & {\bf G}^{\text{K}}\\
{\bf 0}         & {\bf G}^{\text{A}} \\ 
\end{pmatrix},
\end{equation}
with the \emph{retarded}, \emph{advanced} and \emph{Keldysh} components $\textbf{G}^{\text{R,A,K}}$ obeying the relations $\textbf{G}^{\text{A}}=(\textbf{G}^{\text{R}})^{\dagger}$ and $\textbf{G}^{\text{K}} \equiv \textbf{G}^{>} + \textbf{G}^{<}$, $\textbf{G}^{\lessgtr}$ being the \emph{lesser} and \emph{greater} components~\cite{schw.61,keld.65,ra.sm.86,ha.ja}.

\subsection{Dyson equation}\label{sec:Dyson_equation}

The lattice electron GF of central correlated region obeys the Dyson equation~\cite{ti.so.18}
\begin{equation}\label{eq:FullDysonEq}
 \underline{{\bf G}}^{-1}_{zz'}(\omega_n;\epsilon,\overline{\epsilon}) = \underline{{\bf G}}^{-1}_{0,zz'}(\omega_n;\epsilon,\overline{\epsilon}) - \underline{{\bf \Sigma}}_{zz'}(\omega_n;\epsilon,\overline{\epsilon}).
\end{equation}
The GF corresponding to the noninteracting terms in the Hamiltonian~\eqref{eq:Hamiltonian} is 
 \begin{equation}
 \begin{split}\label{eq:non-int_InvGF} 
   & \underline{G}^{-1}_{0,mn,zz'}(\omega_n;\epsilon,\overline{\epsilon}) = \underline{g}^{-1}_{0,mn,zz'}(\omega_n;\epsilon,\overline{\epsilon}) \\ & - \left[ v^2_{\text{l}}\underline{g}_{\text{l}}(\omega_n;\epsilon)\delta_{z,1} + v^2_{\text{r}} \underline{g}_{\text{r}}(\omega_n;\epsilon)\delta_{z,\text{L}} \right] \delta_{mn}\delta_{zz'},
 \end{split}
 \end{equation}
$v_{\text{l}/\text{r}}$ representing the hybridization between leads and correlated region, where we have introduced the shorthand notation $\omega_n \equiv \omega + n\Omega$, $n\in\mathbbm{Z}$. The noninteracting GF of the isolated correlated layer is given by
 \begin{equation}
 \begin{split}\label{eq:inv_non-int_lat_GF_comps}
  & \left[ g^{-1}_{0}(\omega_n;\epsilon,\overline{\epsilon})) \right]^{\text{R}}_{mn,zz'} = \left( \omega_n + \ii 0^{+} -\varepsilon_{z}\right)\delta_{mn}\delta_{zz'} \\ & - t_{zz'}\delta_{mn} - \varepsilon_{mn}(\epsilon,\overline{\epsilon})\delta_{zz'},
  \end{split}
  \end{equation}   
the Keldysh component of which can be neglected because of the leads GF.
 
The internal electric field present in oxide heterostructures, which originates from the polar interfaces between the different correlated layers ~\cite{as.bl.13}, is included via a potential drop $\Phi$ between the outermost layers of the correlated region, i.e. $z=1$ and $z=4$, and is such that the layers' onsite energies vary linearly along the $z$-axis, i.e. 
\begin{equation}\label{eq:on_site_z_dep}
\varepsilon_z =  \varepsilon_z^{(0)} + \frac{\Phi}{2} - \frac{(z-1)\Phi}{(\text{L}-1)},
\end{equation}
where $\varepsilon_z^{(0)} = - U/2$ at half-filling. Such potential drop helps to separate the photoexcited e-h pairs, as shown in Fig.~\ref{fig:energy_setup_final}. 

The Floquet dispersion relation $\varepsilon_{mn}$ for the periodic field in a hyper-cubic lattice is~\cite{ts.ok.08} 
\begin{equation}
\label{eq:Floquet_disp}
\varepsilon_{mn}(\epsilon,\overline{\epsilon}) = 
\begin{cases}
J_{m-n}(\aamp) \ \epsilon & m-n:\text{even} \\
\ii J_{m-n}(\aamp) \ \overline{\epsilon} & m-n:\text{odd},
\end{cases}
\end{equation} 
where $J_n$ denotes the $n$-th order Bessel function of the first kind with the argument $\aamp$ defined as in Sec.~\ref{sec:Model}.

If not otherwise stated, we consider wide-band metallic leads coupled to contact layers with a localized Lorentzian-shaped DOS (see Fig.~\ref{fig:energy_setup_final}), centered at their respective onsite energies $\varepsilon_{\rho}$, to efficiently collect the photoexcited e-h pairs from the central region~\cite{as.bl.13,pe.be.19}.

Following Ref.~\cite{ga.ma.22}, the surface GF $\kel{\vec{g}}_{\rho}$ of the leads reads
\begin{align}\label{eq:bath_GFs}
\begin{split}
g^{\text{R}}_{\rho}(\omega;\epsilon) & = \left( \omega-\varepsilon_{\rho}(\epsilon) +\ii\gamma_{\rho} \right)^{-1} \\
g^{\text{K}}_{\rho}(\omega;\epsilon) & = [g_{\rho}^{\text{R}}(\omega;\epsilon) - g_{\rho}^{\text{A}}(\omega;\epsilon)][1-2f(\omega,\mu_{\rho},\beta)], 
\end{split}
\end{align} 
where $\varepsilon_{\rho}(\epsilon)=\varepsilon_{\rho} + \frac{t_{\rho}}{t^{\ast}_{\parallel}}\epsilon$ denotes the lead dispersion relation and $f(\omega,\mu_{\rho},\beta)=[\ee^{\beta(\omega-\mu_{\rho})}+1]^{-1}$ the Fermi-Dirac distribution function at inverse temperature $\beta$. We recall that the second relation in Eq.~\eqref{eq:bath_GFs} comes from the {\em fluctuation-dissipation theorem}~\cite{st.va.13}.

The electron self-energy (SE) $\underline{\bf{\Sigma}}_{zz'}$ is obtained from real-space Floquet DMFT (F-DMFT) and, in this approximation, is independent of the crystal momentum and spatially local, i.e. $\underline{{\bf \Sigma}}_{zz'}(\omega;\epsilon,\overline{\epsilon}) \simeq \underline{\bf{\Sigma}}_z(\omega)\delta_{zz'}$. Details regarding real-space F-DMFT are given in Sec.~\ref{sec:FDMFT_implementation}.

\subsection{Real-space Floquet DMFT} 
\label{sec:FDMFT_implementation}

The electron SE $\underline{\bf{\Sigma}}_{zz'}$ in Eq.~\eqref{eq:FullDysonEq} is computed using DMFT~\cite{me.vo.89,ge.ko.92,ge.ko.96}, and in particular its nonequilibrium Floquet extension F-DMFT~\cite{ts.ok.08,sc.mo.02u,jo.fr.08}, and real-space generalization for inhomogeneous systems, in which one has to consider the spatial dependence from one coordinate ~\cite{po.no.99,free.04,ec.we.13,okam.07, okam.08,ti.do.16,ti.so.18}. In real-space F-DMFT one considers the electron SE spatially local and neglects its crystal momentum dependence, i.e. $\underline{{\bf \Sigma}}_{zz'}(\omega;\epsilon,\overline{\epsilon}) \simeq \underline{\bf{\Sigma}}_z(\omega)\delta_{zz'}$. Such simplification allows us to solve for each correlated layer $z$ a (nonequilibrium) quantum impurity model with Hubbard interaction $U_z$ and onsite energy $\varepsilon_z$, with a bath hybridization function $\kel{\mat\Delta}_z(\omega)$ determined self-consistently. 

More in detail, the self-consistent real-space F-DMFT scheme is the following: (i) we start from an initial guess for the electron SE $\kel{\mat{\Sigma}}_z(\omega)$, then (ii) we extract the local electron GF as 
\begin{equation}
\label{eq:local_GF}
\kel{\bf G}_{\text{loc},zz}(\omega) = \int \dd\epsilon \int \dd\overline{\epsilon} \ \rho(\epsilon,\overline{\epsilon})\underline{{\bf G}}_{zz}(\omega;\epsilon,\overline{\epsilon}), 
\end{equation}
inverting $\underline{{\bf G}}^{-1}_{zz'}(\omega;\epsilon,\overline{\epsilon})$~\footnote{The outer matrix structure of the inverse lattice GF $\underline{{\bf G}}^{-1}_{zz'}$ is in the layer indices $z,z'$, while the inner one is in the Floquet indices $m,n$, i.e. $\underline{\bf{G}}^{-1}_{zz'}=[({G}^{-1,\textrm{R}/\textrm{K}})_{mn}]_{zz'}$. With the choice of a nearest-neighbor hopping $t_{zz'}$ the overall matrix is tridiagonal, which allows us to apply the \emph{zip algorithm} of Appendix~\ref{sec:zip_alg_FDMFT}.} from Eq.~\eqref{eq:FullDysonEq} and taking the diagonal elements in the \emph{layer} indices. For this goal, either one directly inverts  the matrices in Eq.~\eqref{eq:FullDysonEq}, which results in a major
 computational effort due to the double matrix structure in Floquet and real space, or one uses the recursive Green's function method ~\cite{th.ki.81,free.04,ec.we.13,ti.do.16,le.mu.13}, which we generalize to the Floquet formalism in Appendix~\ref{sec:zip_alg_FDMFT}, which is much faster than the plain matrix inversion. 
(iii) We map the problem onto a single impurity plus bath, with hybridization function 
\begin{equation}\label{eq:imp_Dyson_eq}
\kel{\mat{\Delta}}_z(\omega) = \kel{\mat{g}}^{-1}_{0,z,\text{site}}(\omega) - \kel{\mat{G}}^{-1}_{\text{loc},zz}(\omega) - \kel{\mat{\Sigma}}_z(\omega),
\end{equation}
where $\kel{\mat{g}}^{-1}_{0,z,\text{site}}(\omega)$ is defined as in Eq.~\eqref{eq:inv_non-int_lat_GF_comps} with $\varepsilon_{mn}(\epsilon,\overline{\epsilon})=0$ and $t_{zz'}=0$. (iv) We solve the nonequilibrium many-body impurity problem, which gives the new $\kel{\mat{\Sigma}}_z(\omega)$. (v) The electron SE is inserted into step (ii) and steps (ii)-(v) are iterated until convergence. For the convergence criterion, we explicitly refer to Ref.~\cite{ti.so.18}. 

In Eq.~\eqref{eq:imp_Dyson_eq} for the bath hybridization function $\kel{\mat\Delta}_z(\omega)$, we have a Floquet structure encoding the periodic time dependence. However, in the parameters range we are considering it is safe to adopt the Floquet-diagonal self-energy approximation (FDSA), whereby nondiagonal Floquet indices in $\kel{\mat{\Sigma}}_z(\omega)$ are neglected~\cite{so.do.18,ga.ma.22}. Therefore, the nonequilibrium impurity problem is stationary and we consider only the $(0,0)$-Floquet matrix element of all the quantities in Eq.~\eqref{eq:imp_Dyson_eq}. The other diagonal components of the SE are obtained by exploiting the property $\kel{\Sigma}_{mm}(\omega) = \kel{\Sigma}_{00}(\omega+m\Omega)$. Thanks to the particle-hole inversion symmetry of the system described in Appendix~\eqref{sec:symm_system}, the many-body problem in step (iv) of the real space F-DMFT self-consistent loop has to be solved only for half of the correlated layers. We employ for such goal the auxiliary master equation approach (AMEA)~\cite{ar.kn.13,do.nu.14,do.ga.15,do.so.17,ar.do.18}. We refer to Ref.~\cite{ma.ga.22,ma.we.23,we.lo.23,we.zi.24,ma.we.24} for the details and the latest developments and applications of the AMEA impurity solver.

\subsection{Physical quantities}\label{sec:observables}  

In this section we introduce the observables of interest in this manuscript. We start with the Floquet generalization of the steady-state photocurrent flowing from the left to the right lead~\cite{ti.so.18,okam.07}. Between any of the $(\text{L}+1)$ bonds~\footnote{The bonds include also the one from the left lead ($z=0$) to the first layer ($z=1$) and from the last layer ($z=\text{L}$) to the right lead ($z=\text{L}+1$).} we have the following photocurrent
\begin{align}\label{eq:current_junctions}
 \begin{split}
  j_{\tilde{z},\tilde{z}+1} & = t_{\tilde{z},\tilde{z}+1} \int_{-\Omega/2}^{\Omega/2}\frac{\dd\omega}{2\pi} \int \dd\epsilon \int \dd\overline{\epsilon} \rho(\epsilon,\overline{\epsilon}) \rre\Tr\left[\mat{J}_{\tilde{z},\tilde{z}+1}\right] \\
  & = t_{\tilde{z},\tilde{z}+1} \int_{-\infty}^{+\infty}\frac{\dd\omega}{2\pi} \int \dd\epsilon \int \dd\overline{\epsilon} \rho(\epsilon,\overline{\epsilon}) \rre(J_{00,\tilde{z},\tilde{z}+1})
 \end{split}
\end{align}
where we omitted the frequency and crystal momentum dependence for the sake of simplicity. The new index $\tilde{z}$ takes into account the leads' surface layers {\em and} the correlated region, i.e. $\tilde{z} \in \left\{ 0, 1, \dots,\text{L} \right\}$  with $\tilde{z}=0$ denoting the left and $\tilde{z}+1=\text{L}+1$ the right lead surface, the GF of which are given in Eq.~\eqref{eq:bath_GFs}. The Floquet-represented integrand $\mat{J}_{\tilde{z},\tilde{z}+1}$ in~\eqref{eq:current_junctions} reads
\begin{equation} 
  \mat{J}_{\tilde{z},\tilde{z}+1} = \locGFK{\tilde{z}+1,\tilde{z}} -  \locGFK{\tilde{z},\tilde{z}+1},
\end{equation}
where
\begin{align}\label{eq:GFKoffdiagzzp1}
 \begin{split}
  \locGFK{\tilde{z}+1,\tilde{z}} & = t_{\tilde{z},\tilde{z}+1} \Big [ \locRR{\tilde{z}+1}\locGFK{\tilde{z}} + \locRK{\tilde{z}+1}\locGFA{\tilde{z}} \Big ]  \\
  \locGFK{\tilde{z},\tilde{z}+1} & = t_{\tilde{z},\tilde{z}+1} \Big [ \locLR{\tilde{z}}\locGFK{\tilde{z}+1} + \locLK{\tilde{z}}\locGFA{\tilde{z}+1} \Big ].
 \end{split}
\end{align}
The quantities on the right-hand side of~\eqref{eq:GFKoffdiagzzp1} are described in Appendix~\ref{sec:zip_alg_FDMFT}. In a steady-state situation $j_{\tilde{z},\tilde{z}+1}$ should take on the same value for every $\tilde{z}$. However, due to the limited accuracy of our AMEA impurity solver~\cite{ti.so.18}, deviations occur. For this reason, we introduce the mean $j$ of the photocurrent (averaged over the bonds) and its corresponding standard deviation $\sigma_j$~\footnote{From our simulations, $\sigma_j$ does not decrease improving the convergence accuracy of the DMFT self-consistent procedure.} as  
\begin{align}\label{eq:av_curr_and_std_dev}
 \begin{split}
  j & \equiv \frac{1}{\text{L}+1}\sum_{\tilde{z}=0}^{\text{L}} j_{\tilde{z},\tilde{z}+1} \\
  \sigma_j & \equiv \sqrt{\sum_{\tilde{z}=0}^{\text{L}}\frac{1}{\text{L}+1}(j_{\tilde{z},\tilde{z}+1}-j)^{2}}.
 \end{split} 
\end{align} 
This allows us to provide an estimate for the uncertainty in the photocurrent.
   
In order to characterize the spectral properties of the system at hand we study the \emph{local electron density of states} (DOS) $A_z(\omega)$ and {\em occupation function} $N_{z}(\omega)$~\footnote{The occupation function $N_{z}(\omega)$ indicates how many electronic states per spin per layer are occupied with energy $\omega$ in the layer $z$. The integral over the whole energy range of such quantity gives the number of particles per spin per layer $n_{z}$, i.e. $n_{z}=\int_{-\infty}^{+\infty}\dd\omega N_{z}(\omega)$.} which are defined as   
\begin{equation}\label{eq:el_spectral_function}
 A_z(\omega)=-\frac{1}{\pi}\iim[G_{\text{loc},00,zz}^{\text{R}}(\omega)],
\end{equation} 
and
\begin{equation}\label{eq:spec_occ}
N_z(\omega) = \frac{1}{4\pi} \left\{ \iim \left[G^{\text{K}}_{\text{loc},00,zz}(\omega)\right]-2\iim\left[G^{\text{R}}_{\text{loc},00,zz}(\omega)\right] \right\}.
\end{equation}
In Eqs.~\eqref{eq:el_spectral_function} and~\eqref{eq:spec_occ} $G^{\text{R}}_{\text{loc},00,zz}(\omega)$ represents the time-averaged \emph{retarded} local GF while $G^{\text{K}}_{\text{loc},00,zz}(\omega)$ in Eq.~\eqref{eq:spec_occ} is the corresponding \emph{Keldysh} component. The layer-dependent double occupation 
\begin{equation} \label{eq:double_occ}
N_{\text{D},z} = \langle \hat{n}_{z,\uparrow}\hat{n}_{z,\downarrow} \rangle,  
\end{equation} 
provides a measure of the number of doublons in the upper Hubbard band of the layer $z$. It consists of a \emph{mean-field} contribution $n_{z}^{2}$, which depends on the number of particles per spin per layer $n_{z}=\langle \hat{n}_{z,\uparrow} \rangle = \langle \hat{n}_{z,\downarrow} \rangle$~\footnote{In this work, we consider the central correlated system in a paramagnetic phase.} and is, thus, strongly layer-dependent, and a fluctuation term 
\begin{align}\label{eq:dev_double_occ}
 \begin{split}
  \Delta N_{\text{D},z} 
& = \langle \left( \hat{n}_{z,\uparrow}-\langle \hat{n}_{z,\uparrow} \rangle \right) \left( \hat{n}_{z,\downarrow}-\langle \hat{n}_{z,\downarrow} \rangle \right) \rangle \\   & 
=  N_{\text{D},z}-n_{z}^{2}.
 \end{split}
\end{align} 
This
\emph{fluctuation of the double occupation} $\Delta N_{\text{D},z}$ has a weaker dependence on $z$ and it vanishes for an uncorrelated system at half-filling and is, thus, better suited to discuss the results.
  
\section{Results}\label{sec:results}

\begin{figure*}[t]
\includegraphics[width=0.7\textwidth]{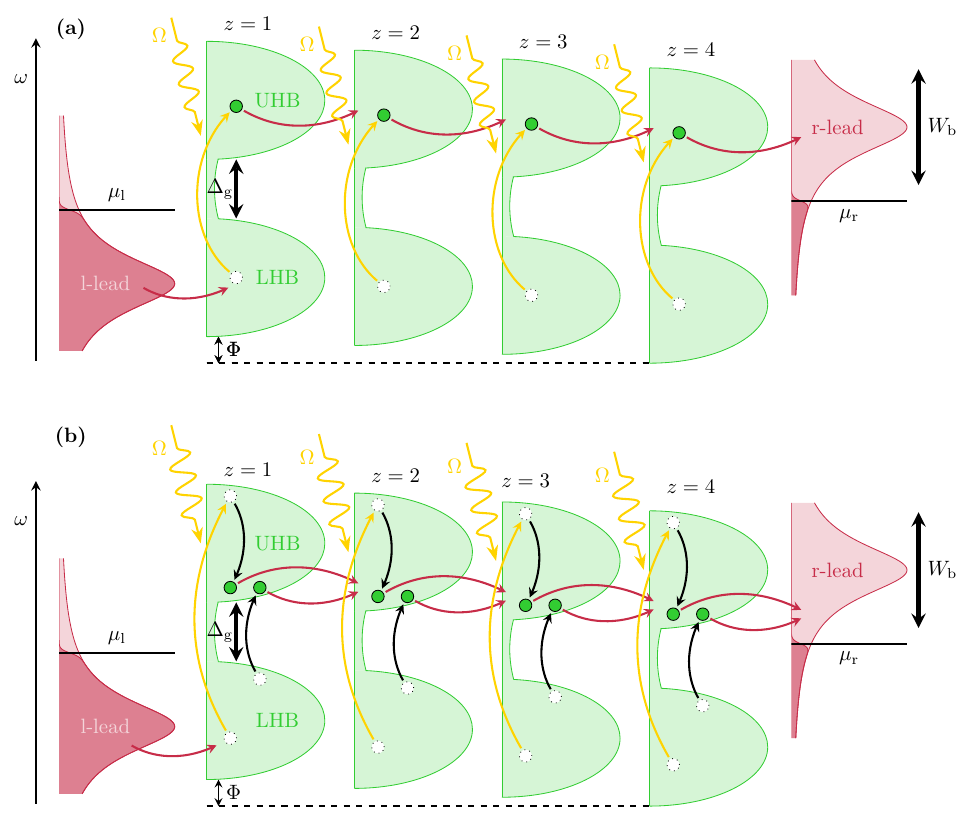}
\caption{Schematic representation of the relevant processes occurring in the system considered in this manuscript. Electrons are injected into the correlated region (green) from the (almost full) left reservoir and drained into the (almost empty) right one (red). The potential drop across the correlated region is denoted as $\Phi$ and is a linear function of the coordinate $z$ of the layers. We set $\mu_r>\mu_l$ so that photovoltaic energy is collected from the device
when current flows from left to right. (a) {\em Direct excitation processes}. When $\Omega < 2\Delta_{\text{g}}$, the photoexcited electron in the upper band escapes directly into the right lead by tunneling through the upper bands. (b) {\em Impact ionization processes}. For large values of $\Omega$, i.e. $\Omega \geq 2\Delta_{\text{g}}$, an electron promoted to the upper band can excite, in turn, a second electron across the gap by transferring energy to it via electron-electron scattering, so that two carriers {\em per photon} can now tunnel towards the right lead.}
\label{fig:energy_setup_final}
\end{figure*}

The setup under investigation is shown in Fig.~\ref{fig:setup_voltmeter_final} while a schematic representation of the energy landscape can be found in Fig.~\ref{fig:energy_setup_final}. We focus on the situation in which the central region consists of $\text{L}=4$ correlated layers, a setup which has already been studied in Refs.~\cite{as.bl.13,pe.be.19}. If not stated otherwise, the default values for the main parameters employed in this work can be found in Table~\ref{tab:default_pars}~\footnote{Given the values of the parameters in Table~\ref{tab:default_pars}, we have that $\alpha\equiv t^{\ast}_{\parallel}E_0/2\Omega^{2}<0.5$, which justifies FDSA~\cite{so.do.18,ga.ma.22}}. 
\begin{table}[b]
  \begin{center}
\begin{tabular}{ cccccccccc }
      \hline
      \hline
        $U$ \ & $E_0$ \ & $\gamma_{\text{l}/\text{r}}$ \ & $v$ \ & $t_{\text{l}/\text{r}}$ \ & $\mu_{\text{l}}$ \ & $t_{\perp}$ \ & $\Phi$ \ & $1/\beta$ \ & $\text{L}$ \\
      \hline
        $12$ \ & $2$ \ & $2$ \ & $0.4$ \ & $2$ \ & $-1$ \ & $1$ \ & $2$ \ & $0.02$ \ & $4$ \\
      \hline
      \hline
    \end{tabular}
    \caption{Default values of the main parameters used in this manuscript: for simplicity we defined $v \equiv v_{\text{l}/\text{r}}$. We recall that the {\em renormalized} hopping in the correlated region is $t^{\ast}_{\parallel}=2$.}
    \label{tab:default_pars}
  \end{center}
\end{table}
The leads' onsite energies $\varepsilon_{\text{l}/\text{r}}$ are chosen such that the left(right) lead DOS overlaps with the lower(upper) Hubbard band of the leftmost(righmost) layer with $\varepsilon_{\text{l}} = \varepsilon_{1}$ and $\varepsilon_{\text{r}}= -\varepsilon_{\text{l}}$, see also the scheme in Fig.~\ref{fig:energy_setup_final}. 
   
The chemical potentials are chosen such that $\mu_{\text{r}} = - \mu_{\text{l}}>0$, so the energy is harvested from periodic driving when an electron current flows
from the left to the right lead, against the chemical potential difference. The gap of the correlated layers is $\Delta_{\text{g}}\approx 3$~\footnote{Due to the hybridization with the leads, a finite spectral weight is present in the gap region of the correlated layers' local DOS, see also Ref.~\cite{ga.ma.22}.}, while the leads' bandwidth is $W_{\text{b}} \approx 8.5$.  
   
In this manuscript our focus is on II processes, in which an electron, promoted to the upper Hubbard band by absorbing the energy of a photon $\Omega$, excites a second electron across the gap by exchanging energy to it via Coulomb interaction.In order for such processes to be energetically allowed,
 the bandwidth of the upper Hubbard band (UHB), which in our case is roughly equal to the leads' DOS, has to be at least twice the size of the gap $\Delta_{\text{g}}$, i.e. $W_{\text{b}} \geq 2\Delta_{\text{g}}$. Only in this case an electron promoted to the UHB has sufficient energy to excite a second one across the gap~\cite{so.do.18,ga.ma.22}.
 
\subsection{Impact ionization}\label{sec:occurrence_II}
   
\begin{figure*}[t]
\includegraphics[width=0.5\textwidth]{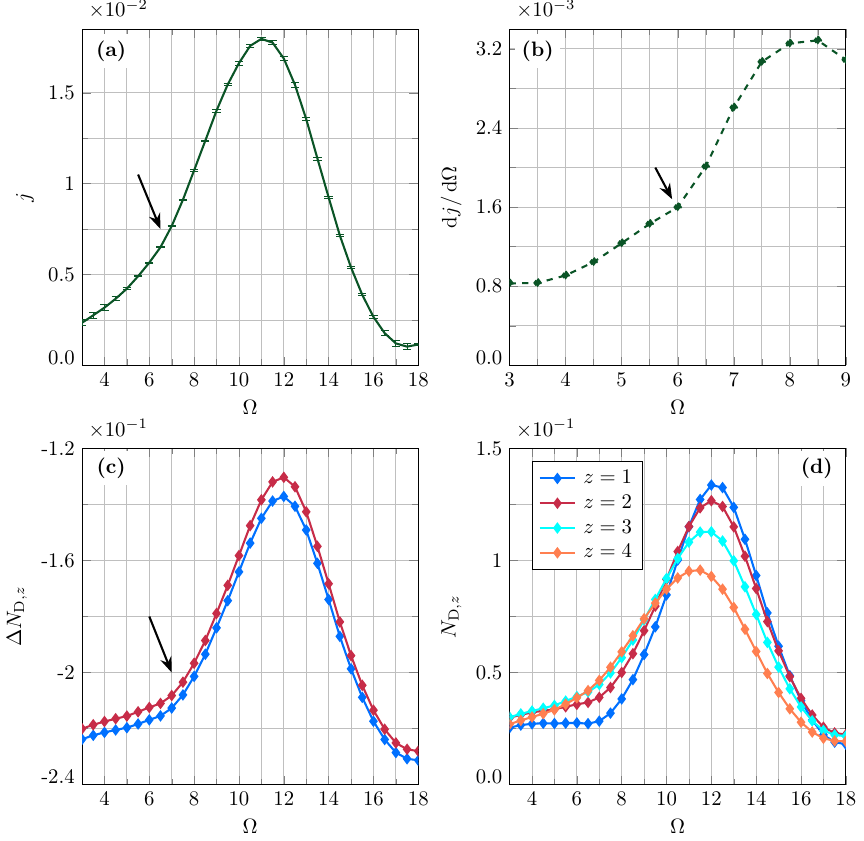}
\caption{(a) Photocurrent $j$ [Eq.~\eqref{eq:av_curr_and_std_dev}] as function of the driving frequency $\Omega$ with error bars corresponding to $\sigma_j$. (b) Numerical derivative of the photocurrent $\text{d}j/\text{d}\Omega$ as function of the driving frequency $\Omega$. (c) Fluctuation of the double occupation $\Delta N_{\text{D},z}$ as function of the driving frequency $\Omega$ for the layers $z=1$ and $z=2$. (d) Double occupation $N_{\text{D},z}$ as a function of the driving frequency $\Omega$ for all the layers of the correlated region. The black arrows in (a)-(b)-(c) highlight the change in slope $\sim 2\Delta_{\textrm{g}}$. Due to the PhI symmetry (see App.~\ref{sec:symm_system}), $\Delta N_{\text{D},\text{L}+1-z} = \Delta N_{\text{D},z}$ so the curves for $z=3$ and $z=4$ in (c) are omitted. Default parameters are specified in Table~\ref{tab:default_pars}. (Here $\Phi=2$, $\mu_{\text{l}}=-1$, $\Delta_{\text{g}} \approx 3$ and $W_{\text{b}} \approx 8.5$.)} 
\label{fig:j_omega_L4U12phi2mu2E4lor_els} 
\end{figure*}
  
The main purpose of this section is to study the onset of the II processes: for this reason we evaluate the photocurrent as function of the driving frequency $\Omega$. As argued in Ref.~\cite{so.do.18,ga.ma.22}, 
the onset of II can be possibly identified by an appreciable change in slope (kink) in the $j$-$\Omega$ curve at $\Omega\sim 2\Delta_{\textrm{g}}$, as it hints at an increased number of carriers being promoted to the UHB, which contribute to a steeper increase of the photocurrent.
 
To better illustrate the point, below we summarize the relevant physical processes that may occur in this periodically-driven system.

\begin{itemize}
\item For $\Omega<\Delta_{\text{g}}$, we expect a negligible photocurrent produced by the residual DOS in the gap, due to the hybridization with the leads.
\item For $\Delta_{\text{g}}<\Omega<\Delta_{\text{g}}+2W_{\text{b}}$ as shown in Fig.~\ref{fig:energy_setup_final}(a), an electron from a lower Hubbard band (LHB) gets photoexcited to the corresponding UHB and moves to the right along the potential drop without further exciting e-h pairs. Such process is often referred to as direct excitation (DE).
\item For $2\Delta_{\text{g}}<\Omega<\Delta_{\text{g}}+2W_{\text{b}}$ as shown in Fig.~\ref{fig:energy_setup_final}(b), a photoexcited electron in the UHB may excite a second one across the gap via II, before both move to the right due to the potential drop.
\item For $\Omega>\Delta_{\text{g}}+2W_{\text{b}}$, a photoexcited electron  ends up near the border of the UHB where the DOS is greatly reduced, and the photocurrent gets strongly suppressed.
 
\end{itemize} 

\subsubsection{Photocurrent and spectral features}\label{sec:photocur_el}

By the analysis of the $j$-$\Omega$ curve in Fig.~\ref{fig:j_omega_L4U12phi2mu2E4lor_els}(a) we see an increase of the photocurrent as $\Omega$ grows larger, with the maximum reached at $\Omega \approx 11$, which is followed by a decrease until $\Omega\approx 17$~\footnote{The current $j$ reaches its minimum at $\Omega\approx 17$ because the \emph{effective current energy window} given by the overlap of the layers' UHBs is smaller than $\Delta_{\text{g}}+2W_{\text{b}}$. Moreover, $j$ does not go to zero as one would expect because of the background current as shown in Ref.~\cite{so.do.18,ga.ma.22}.}. A slight change of slope of the curve around $\Omega\approx7$ seems to indicate an onset of II processes, which are indeed expected to start showing up at $\sim 2\Delta_{\text{g}}$, confirmed by the numerical derivative of $j$ with respect to $\Omega$ in Fig.~\ref{fig:j_omega_L4U12phi2mu2E4lor_els}(b). This is also corroborated by
a similar behavior of the double occupation $N_{\text{D},z}$ as function of $\Omega$ for the different layers $z=\left\{ 1, \dots, 4 \right\}$ displayed in Fig.~\ref{fig:j_omega_L4U12phi2mu2E4lor_els}(d). All these curves show a change in slope around $\Omega\approx7$, although this appears relatively weak for the $z=3$ and $4$ layers. This behavior
is due to the different occupation in the layers as discussed in Sec.~\ref{sec:observables}.
In order to concentrate on electronic correlations effects we plot the corresponding fluctuation $\Delta N_{\text{D},z}$ (see Eq.~\eqref{eq:dev_double_occ})  in Fig.~\ref{fig:j_omega_L4U12phi2mu2E4lor_els}(c) for $z=1$, $2$. These  curves are much less $z$-dependent and, more importantly, display a clear kink at $\Omega \approx 7$, corroborating the onset of II beyond that driving 
frequency.

The $N_{\text{D},z}$ plots display also a reduction of the magnitude and a slight shift of the maximum  with increasing $z$: the reason lies in the (almost) perfect overlap between the rightmost layer's UHB and the right lead's DOS (see Fig.~\ref{fig:energy_setup_final}), which provides electrons with an easy way out into the lead as they tunnel through layers located further to the right. In contrast, $\Delta N_{\text{D},z}$ does not display any layer-dependent magnitude change and position of the maximum. 
   
\begin{figure*}[t]
\includegraphics[width=0.7\textwidth]{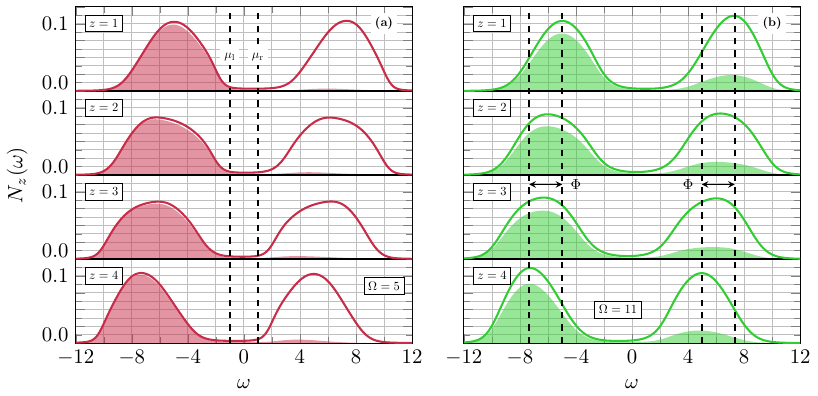}
\caption{Electron DOS $A_z(\omega)$ (solid lines) and occupation function $N_z(\omega)$ (shaded areas) for the different layers at (a) $\Omega=5$ and (b) $\Omega=11$. Dashed vertical lines in (a) mark the position of the leads' chemical potentials $\mu_{\text{l}}$ and $\mu_{\text{r}}$, while those in (b) highlight the separation between the center of the bands of the layers $z=1$ and $z=4$, which roughly equals the potential drop $\Phi$, depicted in Fig.~\ref{fig:energy_setup_final}. Default parameters are specified in Table~\ref{tab:default_pars}. (Here $\Phi=2$, $\mu_{\text{l}}=-1$, $\Delta_{\text{g}} \approx 3$ and $W_{\text{b}} \approx 8.5$.)} 
\label{fig:spec_func_L4U12phi2mu2E4lorOmega5_11}
\end{figure*}

Such indications of II processes are supported by Fig.~\ref{fig:spec_func_L4U12phi2mu2E4lorOmega5_11}, in which the electron DOS $A_{z}(\omega)$ and spectral occupation $N_{z}(\omega)$  are shown for selected values of the driving $\Omega$. As schematically represented in Fig.~\ref{fig:energy_setup_final}, we first notice that the electron DOS of all layers features a finite spectral weight in the gap for both $\Omega=5$ and $\Omega=11$.
At $\Omega=5$, when the photocurrent $j$ is mainly due to DE processes, the UHB of all correlated layers is only very weakly occupied, see Fig.~\ref{fig:spec_func_L4U12phi2mu2E4lorOmega5_11}(a). This signals that the excited electrons are being drained quite effectively out of the central region by the right lead. 
  
At $\Omega=11$, where II is expected to happen, we observe that an important fraction of the electrons now occupies the UHB of all layers in the correlated region, see Fig.~\ref{fig:spec_func_L4U12phi2mu2E4lorOmega5_11}(b). As argued at the beginning of Sec.~\ref{sec:occurrence_II}, a larger occupation of the UHB is consistent with an increased scattering probability among electrons across the gap, the key mechanism for II.

\subsection{Dependence on the electric field amplitude}\label{sec:elec_field_impact}

 In this section we investigate the behavior of the photocurrent and of II upon reducing the amplitude $E_0$ of the driving electric field~\footnote{We point out that the field strength we are considering are rather in the range achieved 
 with intense ultrashort pulsed lasers (see also \cite{mu.we.18}) and far from the intensity of solar radiation. See also the discussion in Sec.~\ref{sec:conclusions}}.

We consider three values of the driving frequency, $\Omega = 5$, $8$ and $11$, in order to  characterize the behavior in the regions before, around and after the onset of II processes.
\begin{figure*}[t]
\includegraphics[width=0.7\textwidth]{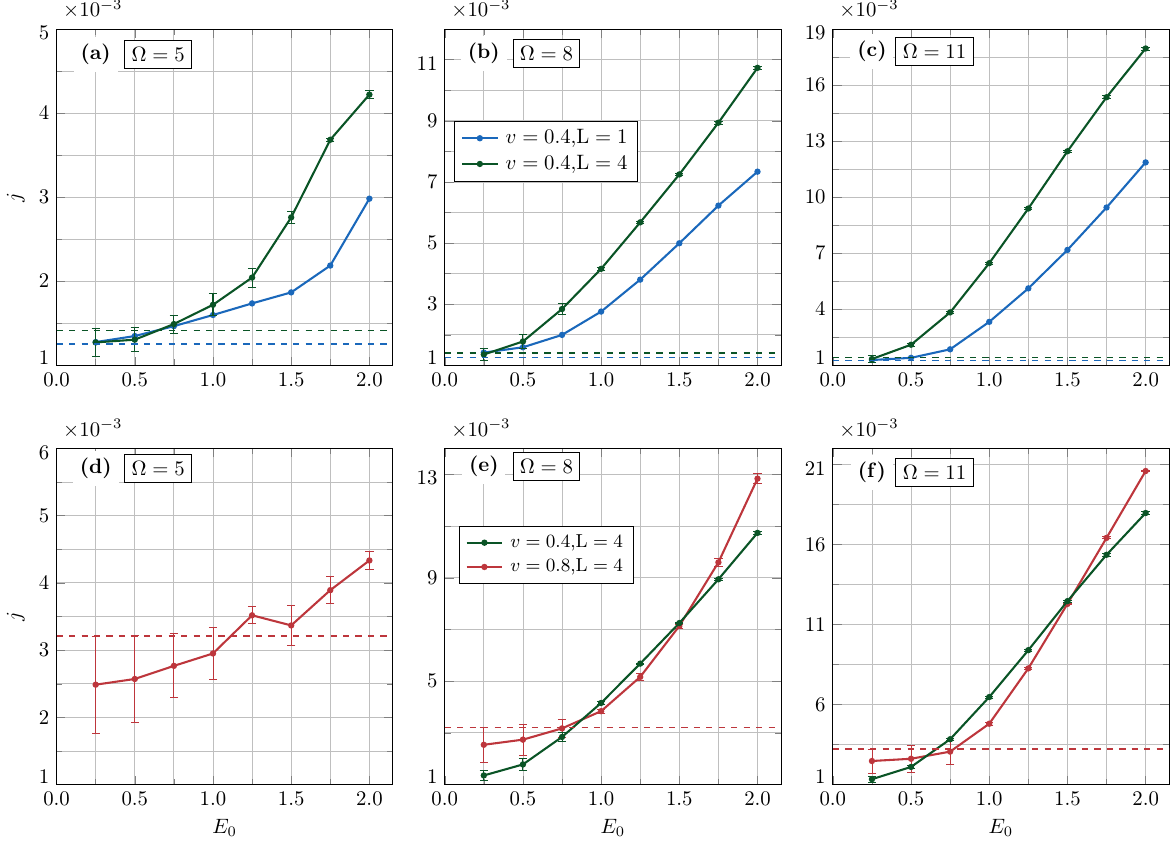}
\caption{Photocurrent $j$ (with error bars $\sigma_j$) as function of $E_0$ at (a)-(d) $\Omega = 5$, (b)-(e) $\Omega = 8$ and (c)-(f) $\Omega = 11$ for different number of correlated layers $\text{L}$ and  hybridizations $v$. Horizontal dashed lines denote the value of the background current obtained with $E_{0}=0$ and refer to the curves of the same color. Green curves in (e)-(f) corresponding to $\text{L}=4$ and $v=0.4$ from (b)-(c) are replotted for comparison. 
Default parameters for the $\text{L}=4$ curves are specified in Table~\ref{tab:default_pars}, while the $\text{L}=1$ curve is obtained with the same parameters used in Ref.~\cite{ga.ma.22}. (Here $\Phi=2$, $\mu_{\text{l}}=-1$, $\Delta_{\text{g}} \approx 3$, $W_{\text{b}} \approx 8.5$)} 
\label{fig:j_E0} 
\end{figure*}  
The $j$-$E_{0}$ curves (in green) displayed in
Fig.~\ref{fig:j_E0}(a)-(c) for $\text{L}=4$ clearly indicate a different behavior of the photocurrent in the cases with and without II. More specifically, in the situation without II ($\Omega=5$, Fig.~\ref{fig:j_E0}(a)) the photocurrent displays a \emph{quadratic} behavior  above the \emph{background} current threshold~\footnote{This \emph{background} current is due to the limited accuracy of our impurity solver~\cite{so.do.18,ga.ma.22}. Below this threshold the photocurrent is inaccurate.} up to electric fields $E_{0}\approx 2$. On the other hand, in the two cases around ($\Omega =8$, Fig.~\ref{fig:j_E0}(b)) and after ($\Omega =11$, Fig.~\ref{fig:j_E0}(c)) the onset of II, the $j$-$E_{0}$ behavior is rather \emph{linear} in a wide region down to $E_0\sim 0.5$.
       
These results suggest a correlation between the occurrence of II and the behavior of the $j$-$E_{0}$ curves. A rough explanation for the deviation from a \emph{quadratic} behavior in the II case may be provided by the following argument, deduced a posteriori from our results in Fig.~\ref{fig:j_E0}. One expects the 
DE photocurrent to be proportional to $|E_{0}|^2$ for small $E_0$, while in presence of II 
$j \propto |E_{0}|^2[1+r_{\textrm{II}}]$, in which the 
II \emph{scattering} rate $r_{\textrm{II}}$~\footnote{The II scattering rate $r_{\textrm{II}}$ is proportional to $U^2$ for small $U$, with a probable saturation for large $U$.} accounts for the creation of two doublons and two holons (see Ref.~\cite{mano.10} for details), increases as function of $\Omega$ and contributes only when $\Omega\gtrsim2\Delta_{\textrm{g}}$. $r_{\textrm{II}}$ is certainly dependent on the occupation of the UHB, and may be responsible for the linear behavior of the photocurrent above a certain threshold field $E\geq E_{0,\textrm{th}}$. We  argue that this crossover from a quadratic to a linear-like $j$-$E_{0}$ behavior may be considered as a signature of II.
 

The value of $E_{0,\textrm{th}}$, above which the curve $j$-$E_{0}$ becomes linear, increases when decreasing the number of layers $L$ in the central region, as one sees from Fig.~\ref{fig:j_E0}(a)-(c), which  displays the $j$-$E_{0}$ curves
for $\text{L}=1$~\footnote{The data for the $\text{L}=1$ case are obtained with the same parameters used in Ref.~\cite{ga.ma.22}.}. This can be understood by the observation that excited doublons exhibit a swifter departure from a more confined correlated region, which in turn decreases the probability of II for a fixed $E_{0}$. A similar effect is observed when increasing the hybridization to the leads as displayed in Fig.~\ref{fig:j_E0}(e)-(f), as already observed in Ref.~\cite{ga.ma.22}. Due to the larger value $v=0.8$~\footnote{The data for the $v=0.8$ are obtained with the default parameters specified in Table~\ref{tab:default_pars}.} which disfavors II, the linear regime sets in at larger field amplitudes ($E_0 \gtrsim 1$) and it is more marked at larger driving frequencies ($\Omega=11$) with respect to $v=0.4$, see Fig.~\ref{fig:j_E0}(e)-(f)).  
 
\begin{figure*}[t]
\includegraphics[width=0.7\textwidth]{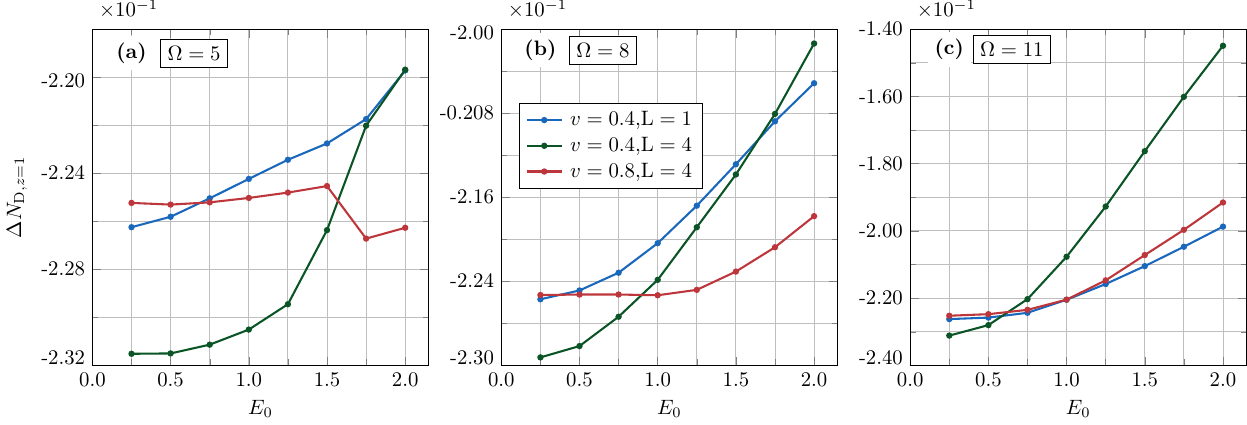}
\caption{Fluctuation of double occupancy $\Delta N_{\text{D},z}$ for the layer $z=1$ as function of $E_0$ at (a) $\Omega = 5$, (b) $\Omega = 8$ and (c) $\Omega = 11$ for different number of correlated layers $\text{L}$ and  hybridizations $v$. Default parameters for the $\text{L}=4$ curves are specified in Table~\ref{tab:default_pars}, while the $\text{L}=1$ curve is obtained with the same parameters used in Ref.~\cite{ga.ma.22}. (Here $\Phi=2$, $\mu_{\text{l}}=-1$, $\Delta_{\text{g}} \approx 3$, $W_{\text{b}} \approx 8.5$)}
\label{fig:deviation_double_occ_E0} 
\end{figure*}

Comparing the green and red curves in Fig.~\ref{fig:j_E0}(e)-(f) we further notice that the photocurrent for $v=0.4$ is larger than for $v=0.8$ in an intermediate region of $E_0$-values, which becomes wider for $\Omega=11$. The reason is the following: if the 
field amplitude is large, the number of photoexcited carriers in the UHB will be also consistenly large and a higher hybridization will provide a larger contribution to the
photocurrent. On the other hand, for weaker fields the occupations of the upper bands is sensibly smaller and a carrier multiplication mechanism as II becomes more important. 
\begin{figure*}[t]
\includegraphics[width=0.7\linewidth]{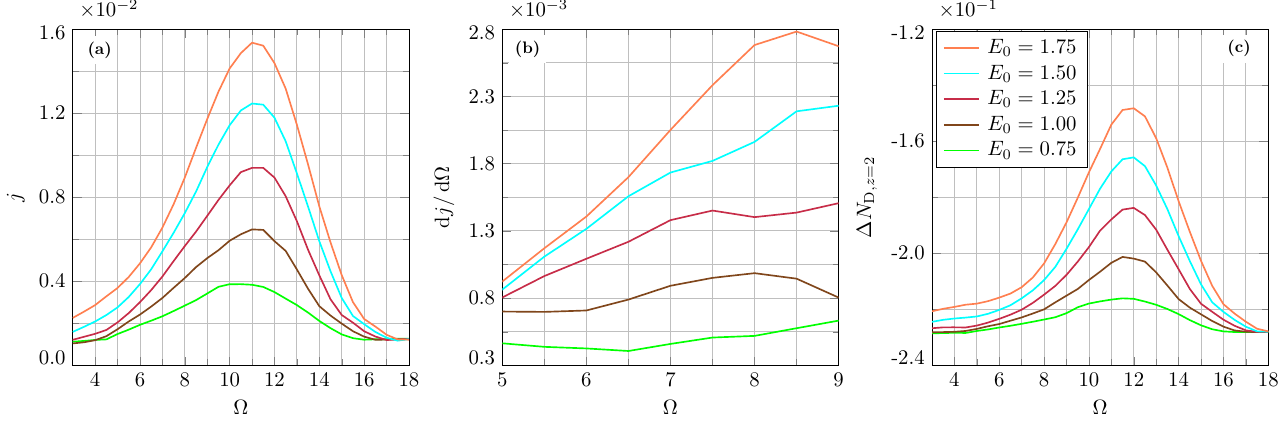}
\caption{(a) Photocurrent $j$, (b) numerical derivative of the photocurrent $\text{d}j/\text{d}\Omega$ and (c) fluctuation of double occupancy $\Delta N_{\text{D},z}$ for the layer $z=2$, as function of the driving frequency $\Omega$ for selected values of the electric field amplitude $E_0$. Error bars corresponding to $\sigma_j$ in (a) are not shown for better visualization of the data. Default parameters are specified in Table~\ref{tab:default_pars}. (Here $\Phi=2$, $\mu_{\text{l}}=-1$, $\Delta_{\text{g}} \approx 3$ and $W_{\text{b}} \approx 8.5$.)}
\label{fig:j_omega_L4U12phi2mu2E4lorsmallerE} 
\end{figure*}
To complete our analysis, in Fig.~\ref{fig:j_omega_L4U12phi2mu2E4lorsmallerE} we analyze the behavior of the $j$-$\Omega$ curves for values of the field amplitude $E_0=\left\{ 0.75, \dots, 1.75 \right\}$. These curves display no appreciable change in slope for $E_0\lesssim 1.25$ and only a slight bend for larger fields (see Fig.~\ref{fig:j_omega_L4U12phi2mu2E4lorsmallerE}(a)) around $\Omega\approx5-7$, even though II 
is expected  to occur at $E_0\gtrsim 0.5$, as suggested by  Fig.~\ref{fig:j_E0}(b)-(c)). 
On the other hand, the numerical derivative of the photocurrent $\text{d}j/\text{d}\Omega$ in Fig.~\ref{fig:j_omega_L4U12phi2mu2E4lorsmallerE}(b) increases for $\Omega\sim 2\Delta_{\textrm{g}}$ and the $\Delta N_{\text{D}}$-$\Omega$ curve in Fig.~\ref{fig:j_omega_L4U12phi2mu2E4lorsmallerE}(c) displays a small but clearer kink already at smaller field amplitudes, in support of II. The behavior of this quantity
 as function of $E_0$ in Fig.~\ref{fig:deviation_double_occ_E0} confirms the II onset from $E_0\gtrsim 0.5$, with a clear transition from the small $E_0$ quadratic regime to a linear one for $\Omega=8,11$ (green curve) in accordance with Fig.~\ref{fig:j_E0}. Also the curves for $L=1$ and $v=0.8$ in Fig.~\ref{fig:deviation_double_occ_E0}(b)-(c) undergo a crossover to a linear behavior at larger $E_0$-values ($E_0\gtrsim 1$), here especially evident for $\Omega=11$, as already observed in Fig.~\ref{fig:j_E0}.

\subsection{Direct connection to wide-band metallic leads}\label{sec:WBL}
               
In contrast to the setup of Fig.~\ref{fig:energy_setup_final} we now consider the case of a direct connection between the metallic wide-band leads and the correlated layers without intermediate narrow-band contacts. It's interesting to study this situation because the carriers which are photoexcited to the upper Hubbard band may now escape to the "wrong" side (here the left lead) of the device and only the potential gradient 
induces a net flow of charges moving in the "right" direction (the right lead). 

By taking the metallic leads in the wide-band limit (WBL), 
the \emph{retarded} component of the leads' surface GF in Eq.~\eqref{eq:bath_GFs} reads
\begin{equation}\label{eq:WBL_bathGF}
v_{\rho}^{2}g^{\text{R}}_{{\text{b},\rho}}(\omega, {\bf k}) \approx - \frac{i}{2} \ \Gamma_{\rho},
\end{equation}
where~\footnote{Formally the WBL corresponds to taking 
$v_{\rho}\to\infty$, $\gamma_{\rho}\to\infty$ with finite $\Gamma_{\rho}$.}
 $\Gamma_{\rho} \equiv 2v_{\rho}^{2}/\gamma_{\rho}$, while the \emph{Keldysh} component in~\eqref{eq:WBL_bathGF} is obtained by means of the \emph{fluctuation-dissipation} theorem~\cite{st.va.13}. The WBL parameters are such that the magnitude of the gap and the bandwidth of the Hubbard bands are the same as in Sec.~\ref{sec:results}, i.e. $\Delta_{\text{g}} \approx 3$ and $W_{\text{b}} \approx 8.5$. 
\begin{figure*}[t]
\includegraphics[width=0.9\textwidth]{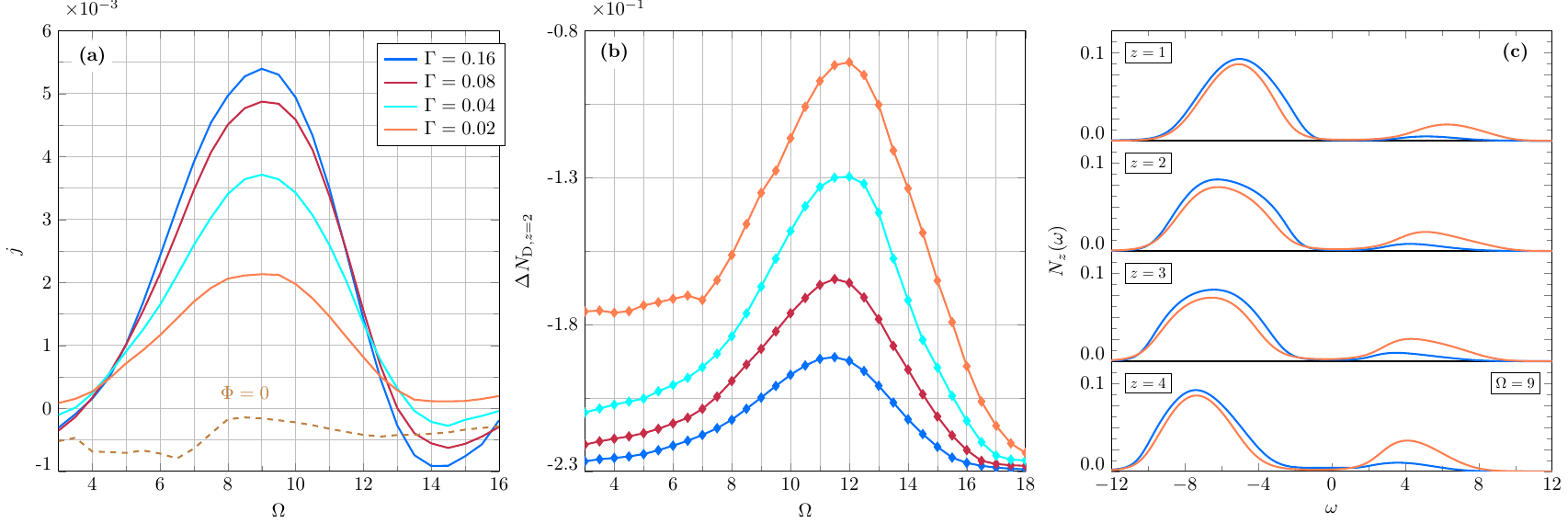}
\caption{(a) Photocurrent $j$ and (b) fluctuation of double occupancy $\Delta N_{\text{D},z}$ for the layer $z=2$, as function of the driving frequency $\Omega$ for selected values of the WBL rate $\Gamma$. (c) Occupation function $N_z(\omega)$ for the different layers at $\Omega=9$ for $\Gamma=0.16,0.02$. Dashed brown line in (a) represents the photocurrent $j$ for $\phi=0$ at $\Gamma=0.16$. Error bars corresponding to $\sigma_j$ in (a) are not shown for better visualization of the data. Default parameters are specified in Table~\ref{tab:default_pars}. (Here $\Phi=2$, $\mu_{\text{l}}=-1$, $\Delta_{\text{g}} \approx 3$ and $W_{\text{b}} \approx 8.5$.)}
\label{fig:WBL_diff_Gammas} 
\end{figure*}

We start by analyzing the photocurrent $j$ as function of the driving frequency $\Omega$ for different values of the WBL rate $\Gamma=\Gamma_{\rho}$. In Fig.~\ref{fig:WBL_diff_Gammas}(a)~\footnote{The curves relative to the $\Gamma=0.02$ for $\Omega<7$ exhibit some numerical instabilities due to the very small coupling to the leads.} we observe that the peak of the photocurrent $j$ occurs at smaller driving frequencies $\Omega\approx 9$ and 
 its magnitude is roughly one order of magnitude smaller than the results in Fig.~\ref{fig:j_omega_L4U12phi2mu2E4lor_els}(a)~\footnote{Larger values of the driving frequency, i.e. $\Omega \gtrsim14$, are responsible for a \emph{non-physical} increase of $j$.}. This is due to the mentioned effect that  photoexcited carriers can escape to the "wrong" side of the device. Moreover, the slight kink $\sim 2\Delta_{\textrm{g}}$ in the $j$-$\Omega$ curve present in Fig.~\ref{fig:j_omega_L4U12phi2mu2E4lor_els}(a) is here completely absent at the corresponding $\Gamma=0.16$ (blue curve in Fig.~\ref{fig:WBL_diff_Gammas}(a)). For smaller values of the WBL rate $\Gamma$ there is also no evidence of such change of slope, which instead would be present with lower hybridizations in the setup discussed in Secs.~\ref{sec:occurrence_II} and~\ref{sec:elec_field_impact}, due to the presence of II. Considering also the large field amplitude $E_0=2$ chosen in this situation, this clearly points to a drastic reduction of II in this setup.
 
 On the other hand, a kink around $\sim 2\Delta_{\textrm{g}}$ can be observed in $\Delta N_{\text{D},z}$, see Fig.~\ref{fig:WBL_diff_Gammas}(b). This kink gets particularly pronounced for small $\Gamma$, and is connected with a higher occupation of the UHB of the heterostructure (Fig.~\ref{fig:WBL_diff_Gammas}(c)). This may favor II, although its signatures are absent in the photocurrent.
  
Without potential gradient ($\Phi=0$), the photocurrent flows in the opposite direction (right to left), as shown in 
Fig.~\ref{fig:WBL_diff_Gammas}(a). This is expected, as the e-h pairs created by photoexcitations don't have a preferred direction and the drag of charges from the left lead equals the flow to the right so that the chemical potential imbalance determines the net current.

Summarizing, such WBL analysis indicates
that a setup with narrow-band intermediate contacts, as interfacial Ti-layers in Refs.~\cite{as.bl.13,pe.be.19},
is much more favorable for II than a direct coupling to wide-band metallic leads.
  
\section{Conclusions}
\label{sec:conclusions}
 
In this manuscript we study a simplified setup for a Mott photovoltaic system consisting of $\text{L}=4$ correlated layers displaced along the $z$-axis under the driving of a monochromatic, periodic electromagnetic field with frequency $\Omega$. The correlated layers are connected to wide-band metallic leads at different chemical potentials with a narrow-band layer in between, so to reach a (Floquet) steady-state photocurrent and collect the energy of the electromagnetic field. Correlations are treated within Floquet dynamical mean-field theory with an accurate steady-state impurity solver.

We find that impact ionization contributes substantially to the photocurrent for driving frequencies larger than twice the gap at the considered electric field amplitudes.
Our results further suggest a correlation between impact ionization and the onset of a linear behavior of the photocurrent as a function of the field strength. 
Impact ionization is amplified by increasing the thickness of the correlated region and/or reducing the hybridization with the metallic leads. This enhancement occurs because these factors increase the "duration" of the stay of photoexcited carriers in the upper Hubbard bands. As a result, there is a larger probability for additional scattering events, which consequently increases the likelihood of generating extra electron-hole pairs. 

On the other hand, a direct coupling to wide-band metallic leads without intermediate narrow-band layers reduces the photocurrent by an order of magnitude and drastically suppresses the contribution of II. This is consistent with the choices made in Refs.~\cite{as.bl.13,pe.be.19}: in order to efficiently collect photoexcited carriers produced by the external driving and obtain a higher efficiency in Mott photovoltaic setups, one has to narrow the wide band of the metallic leads so as to achieve a localized density of states on the surface, which matches the LHB of the first and the UHB of the last correlated layer of the heterostructure. This enables to inject and drag particles in and out of selected regions of the energy spectrum.  

Some comments are in order. The electromagnetic field amplitudes and intensities of the external driving discussed in this manuscript are orders of magnitude larger than those typical of the sunlight and are usually achievable only in experiments with ultrashort pulsed lasers~\cite{mu.we.18}. On the other hand, a thicker Mott photovoltaic device would consist of a larger number of layers $\text{L}$, and as demonstrated, the threshold field $E_{0,\textrm{th}}$ decreases with increasing $\text{L}$. While the current analysis does not give a definitive assessment of the rate of this decrease, it cannot then be excluded that for larger devices, II may occur at intensities comparable to those provided by solar radiation. Addressing such possibility would be an interesting question for future investigations.
 
Furthermore, it is well-known~\cite{free.04,okam.07,okam.08} that a central role for transport properties in strongly-correlated heterostructures is played by the overlap of the density of states of the different layers. When phonons are taken into account, electronic scattering processes and spectral features are inevitably modified, which reflect in  changes of
 the photocurrent. Multiple orbitals and magnetic effects also affect drastically the layers' density of states and will change considerably the transport properties as already shown in Ref.~\cite{ec.we.14,pe.be.19}. For future studies then, the inclusion of such effects  will be an important step to better describe and understand the mechanisms of impact ionization in a more realistic setup and to provide useful benchmarks for possible experimental realizations. 
      
\begin{acknowledgments} 
We thank D. Golez, Y. Murakami, F. Petocchi, P. Werner and T. Jauk for fruitful discussions. This work was supported by the Austrian Science Fund (FWF) within Project P 33165-N, as well as NaWi Graz. E. Arrigoni conceived the project, supervised the work and helped in several technical aspects. P. Gazzaneo designed and implemented the multilayer scheme and produced theoretical data. D. Werner contributed the impurity solver. T. M. Mazzocchi contributed the first drafts of this manuscript. The manuscript has been drafted by P. Gazzaneo with contributions from all authors. The computational results presented have been obtained using the Vienna Scientific Cluster and the L-Cluster Graz.
\end{acknowledgments} 

\appendix  

\section{Generalization of the \emph{zip algorithm} for real-space F-DMFT}
\label{sec:zip_alg_FDMFT}
 
We present here the generalization to the Floquet formalism of the recursive Green's function method \cite{free.04,ec.we.13,ti.do.16}, also known as \emph{zip algorithm}. To facilitate the reading, we omit the $(\omega_n,\epsilon,\overline{\epsilon})$-dependence.

In this procedure, the expressions for the lattice GF in Eq.~\eqref{eq:FullDysonEq} are
\begin{align}
\textbf{G}^{\text{R}}_{zz} &= \big( \locGFRinv{zz} - t_{z-1,z}^2 \locLR{z-1} - t_{z,z+1}^2 \locRR{z+1} \big)^{-1} \label{eq:GzzR}, \\ 
\textbf{G}^{\text{K}}_{zz} &= - \textbf{G}^{\text{R}}_{zz}\locGFKinv{z}\textbf{G}^{\text{A}}_{zz}\label{eq:GzK},
\end{align} 
with $z\in \left\{ 1,...,\text{L} \right\}$, where 
\begin{align}
\locGFKinv{z} &= [{\bf g}^{-1}]^\text{K}_{z} - t_{z-1,z}^2 \locLK{z-1} - t_{z,z+1}^2 \locRK{z+1}\label{eq:GinvzK}, \\
 [g^{-1}]^\text{K}_{mn,z} &= -\left[ v^2_{\text{l}}g_{\text{l}}^\text{K}\delta_{z,1} + v^2_{\text{r}} g_{\text{r}}^\text{K}\delta_{z,\text{L}} \right]\delta_{mn}\notag \\ &- \Sigma^\text{K}_{mn,z} \label{eq:GF_isol_keld_zip},
\end{align} 
 $\underline{{\bf L}}_{0}=0$ and $\underline{{\bf R}}_{L+1}=0$. We denote with $\underline{{\bf L}}_{z-1}$ and $\underline{{\bf R}}_{z+1}$ respectively the GFs for the $z-1$-th and $z+1$-th layer of the isolated system at the left and at the right of the $z$-th layer, which we call from now the \emph{left} and \emph{right} GF. The boldface stays for a Floquet-represented matrix, as in Sec.~\ref{sec:Method_formalism}. As a remark, the inverse lattice \emph{Keldysh} GFs in Eqs.~\eqref{eq:GinvzK} and~\eqref{eq:GF_isol_keld_zip} are diagonal in the \emph{layer} indices because of the real-space DMFT approximation for the electron SE, i.e. $\underline{{\bf \Sigma}}_{zz'} \simeq \underline{\bf{\Sigma}}_z\delta_{zz'}$.   
  
The \emph{left} GF is obtained recursively as follows:
\begin{align}
  \locLR{z} &= \big(\locGFRinv{zz} - t_{z-1,z}^2 \locLR{z-1} \big)^{-1} \label{eq:LzR}, \\
 \locLK{z} &= - \locLR{z}\big([{\bf g}^{-1}]^\text{K}_{z}-t_{z-1,z}^2\locLK{z-1}\big)\locLA{z}\label{eq:LzK},
\end{align} 
for $z\in \left\{ 2,...,\text{L-1} \right\}$, with the initial conditions 
\begin{align}
\locLR{1} &= \textbf{G}^{\text{R}}_{11}\label{eq:L1R}, \\
\locLK{1} &= - \locLR{1}[{\bf g}^{-1}]^\text{K}_{1}\locLA{1} \label{eq:L1K}.
\end{align}
In the same way for the \emph{right} GF:
\begin{align}
  \locRR{z} &= \big( \locGFRinv{zz} - t_{z+1,z}^2 \locRR{z+1} \big)^{-1}\label{eq:RzR}, \\
  \locRK{z} &= - \locRR{z}\big([{\bf g}^{-1}]^\text{K}_{z}-t_{z,z+1}^2\locRK{z+1}\big)\locRA{z} \label{eq:RzK},
\end{align} 
for $z\in \left\{ \text{L-1},...,2 \right\}$, with the initial conditions
\begin{align}
 \locRR{\text{L}} &= \textbf{G}^{\text{R}}_{\text{LL}} \label{eq:RLR}, \\
  \locRK{\text{L}} &= - \locRR{\text{L}}[{\bf g}^{-1}]^\text{K}_{\text{L}}\locRA{\text{L}} \label{eq:RLK}.
\end{align}

\section{Particle-hole inversion symmetry}
\label{sec:symm_system}

The system, with the Hamiltonian given in the main text in Eq.~\eqref{eq:Hamiltonian}, the choice of the onsite energies in Sec.~\ref{sec:Dyson_equation} and the parameters chosen as in Sec.~\ref{sec:results}, is invariant under a simultaneous \emph{particle-hole} transformation and reflection of the $z$-axis respect to the centre of the heterostructure \cite{ti.do.16,ti.so.18}. For this reason, the properties of $z$-th and ${(\text{L}+1-z)}$-th layers are related by a \emph{particle-hole} transformation. Due to this PhI (\emph{particle-hole inversion}) symmetry, the $(0,0)$-Floquet matrix element of the SEs obeys the relations:   
\begin{align}
\Sigma_{\text{L}+1-z,00}^{\textrm{R}}(\omega)&=-[\Sigma_{z,00}^{\textrm{R}}(-\omega)]^*+U_z,\label{eq:mirror_symm_el_SE_ret} \\
\Sigma_{\text{L}+1-z,00}^{\textrm{K}}(\omega)&=[\Sigma_{z,00}^{\textrm{K}}(-\omega)]^*\label{eq:mirror_symm_el_SE_kel}. 
\end{align}
The other diagonal Floquet elements of the SEs are reconstructed by using $\kel{\Sigma}_{mm}(\omega) = \kel{\Sigma}_{00}(\omega+m\Omega)$. Thanks to the relations in Eqs.~\eqref{eq:mirror_symm_el_SE_ret} and ~\eqref{eq:mirror_symm_el_SE_kel}, one has to solve in the real-space F-DMFT loop only the many-body impurity problems relative to half of the correlated region.  
 
For the quantities defined in Sec.~\ref{sec:observables} the PhI symmetry gives: 
\begin{align}
 A_{\text{L}+1-z}(\omega) &= A_z(-\omega)\label{eq:mirror_symm_el_DOS}, \\
n_{\text{L}+1-z} &= 1-n_{z} \label{eq:mirror_symm_nz}, \\ 
N_{\text{D},\text{L}+1-z} &= 1-2n_{z}+ N_{\text{D},z} \label{eq:mirror_symm_double_occ}, \\
\Delta N_{\text{D},\text{L}+1-z} &= \Delta N_{\text{D},z} \label{eq:mirror_symm_dev_double_occ}. 
\end{align}
%

    
\bibliography{reference_database_copied}
    
\end{document}